\documentclass[preprint,aps]{revtex4} 

\usepackage{epsf}
\usepackage{color}

\begin{document}

\title{Quantum $s=1/2$ Antiferromagnets on Archimedean Lattices: The Route from Semiclassical Magnetic Order to Nonmagnetic Quantum States}

\author{D. J. J. Farnell}

\affiliation{Division of Mathematics and Statistics, Faculty of Advanced Technology, 
University of  South Wales, Pontypridd CF37 1DL, Wales, United Kingdom}
\author
{
O. G\"otze and J. Richter 
}
\affiliation
{
Institut f\"ur Theoretische Physik, Universit\"at Magdeburg, D-39016 Magdeburg, Germany
}
\author
{
R. F. Bishop and P. H. Y. Li
}
\affiliation
{
School of Physics and Astronomy, The University of Manchester, Schuster Building, Manchester M13 9PL, United Kingdom
}

\date{\today}

\begin{abstract}
We investigate ground states of $s$=1/2 Heisenberg antiferromagnets
on the eleven two-dimensional (2D) Archimedian lattices by using the 
coupled cluster method.
Magnetic interactions and quantum fluctuations play against 
each other subtly in 2D quantum magnets and the magnetic 
ordering is thus sensitive to the features of lattice topology. 
Archimedean lattices are those lattices that have
2D arrangements of regular polygons and they often build the
underlying magnetic lattices of insulating
quasi-two-dimensional quantum magnetic materials.
Hence they allow a systematic study of the relationship between 
lattice topology and magnetic ordering. 
We find  that the Archimedian
lattices fall into three groups:   
those with semiclassical
magnetic ground-state long-range order, those with a magnetically  disordered
(cooperative quantum paramagnetic) ground state, and those 
with a fragile magnetic order.
The most relevant parameters affecting the magnetic ordering  
are the coordination number
and the degree of frustration present.   
\end{abstract}
\maketitle

\begin{figure}[ht]
\epsfxsize=13cm
\centerline{\epsffile{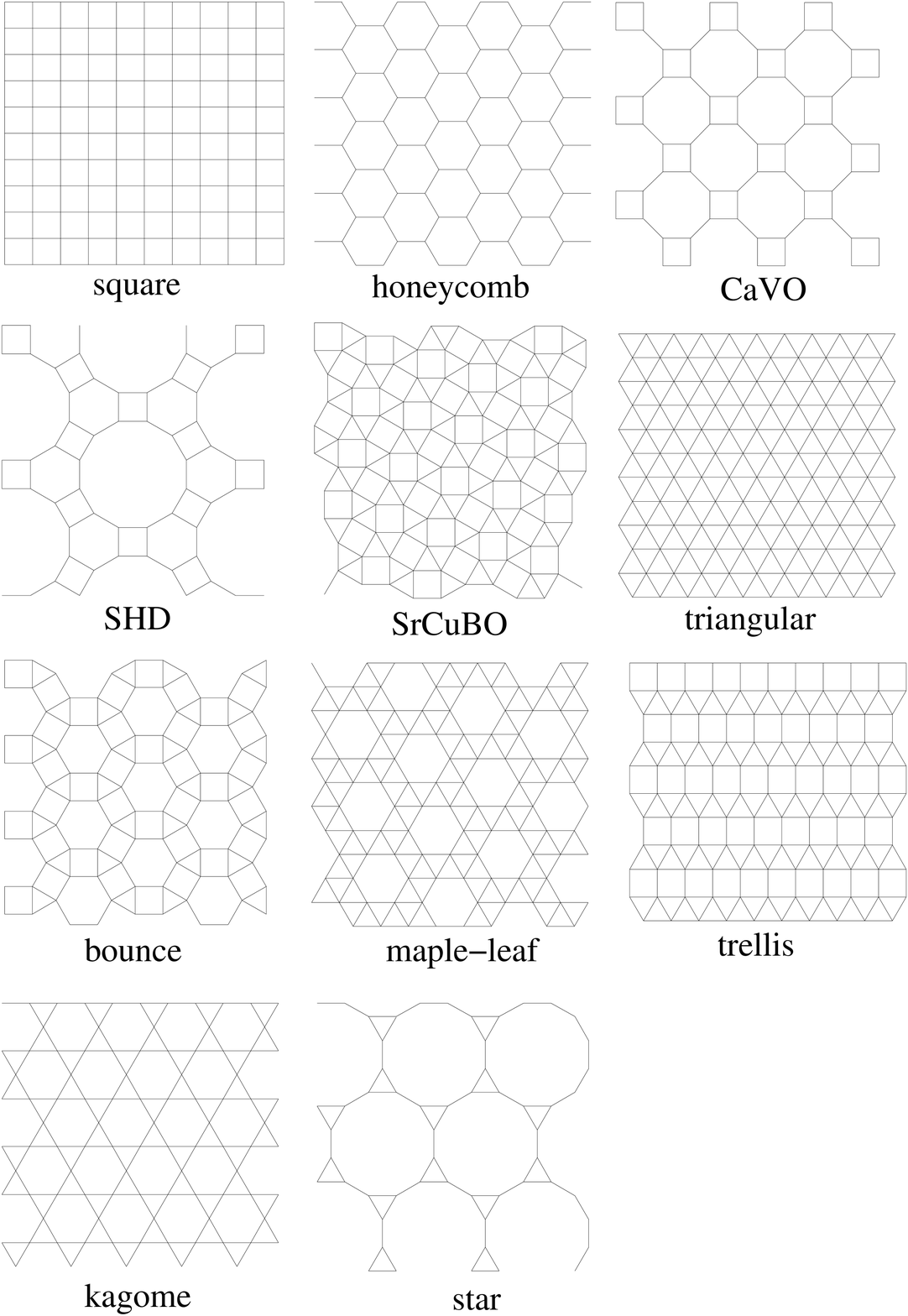}}
\caption{The eleven Archimedean lattices.}
\label{fig1}
\end{figure}

In two-dimensional (2D) quantum Heisenberg antiferromagnets (HAFMs)
the balance  between quantum
fluctuations and interactions  depends subtly on the topology of the underlying
lattice. Thus, a large variety of ground state (GS) phases are found in
2D quantum magnets, among them exotic quantum states, see,
e.g., \cite{balents,sachdev}.
The prototypes of 2D arrangements of spins
are the 11 uniform Archimedean lattices (ALs), see, e.g.,
\cite{gruenbaum,ed_archimedean}, which present an ideal
playground
for a systematic study of the interplay between 
lattice topology, magnetic interactions and quantum fluctuations.
ALs are formed from 2D arrangements of 
regular polygons. Moreover, all sites of
a certain  AL are topologically equivalent, but the nearest-neighbor (NN) bonds are allowed to
be topologically inequivalent.  Well-known (and well
studied) members of the ALs are the  square, honeycomb, triangular,  
and kagome
lattices.  More exotic (and less studied) lattices are the star, ``CaVO'',
``SHD'', maple-leaf, trellis, ``SrCuBO'' and bounce lattices, see, e.g., 
Fig.~\ref{fig1}.

Four of the ALs (namely square, honeycomb, CaVO, and SHD) 
are bipartite lattices (i.e. only even polygons are
present). In the other seven ALs triangular polygons are present and 
the HAFM is frustrated. 
In particular, the triangular and the kagome lattices have attracted much
attention as paradigms of 2D frustrated lattices, see, e.g.,
Refs.~\onlinecite{dmrg_trian,zhito2009,tanaka,mendels,Yan2011,lauchli2011,scholl,kagome_general_s,normand2013}. 
Interestingly, not only the well-known ALs are found to be underlying lattice
structures of the magnetic ions of various compounds, but also the
more exotic ones are realized, see, e.g., 
CaV$_4$O$_9$ (CaVO) \cite{Tan},  
SrCu$_2$(BO$_3$)$_2$ (SrCuBO) \cite{Kag}, 
a polymeric iron(III) acetate (star)\cite{star} or 
$\mathrm{M_x}[\mathrm{Fe}(\mathrm{O}_2\mathrm{CCH}_2)_2\mathrm{NCH}_2\mathrm{PO}_3]_6\cdot
\mathrm{n H}_2\mathrm{O}$ and Cu$_6$Al(SO$_4$)(OH)$_{12}$Cl$\cdot$3H$_2$O
(maple-leaf)\cite{cave,fennell}.
Very recently, an overview of the experimental realizations 
of Archimedean spin  lattice materials (and from the point-of-view of a chemist) 
has been given in Ref. [\onlinecite{Zheng}].
Hence, a systematic and comparative investigation of 
the HAFM on the ALs is not only interesting as a ``paradigmatic'' study of
the role of topology in 2D
quantum systems but also
from the experimental point of view in the field of quantum magnetism.
Let us also mention here, that the special lattice topology of the ALs plays a role
in a large variety of interacting quantum system such as Chern
insulators, see e.g. Refs.~\onlinecite{fiete2011,bernevig2012}, or  chiral spin
liquids, see e.g. Ref.~\onlinecite{kivelson2007}.

A first attempt to study the GS properties systematically was given in
Ref.~\onlinecite{ed_archimedean} where exact diagonalization (ED) data for
the  GS energies and order parameters for 
the spin-1/2 HAFM on the ALs
were presented.
ED is severely limited by the maximum 
lattice size that can be treated by using even very 
large computational resources \cite{starI,ED40,lauchli2011}. Since only two of the ALs are primitive 
lattices with only one site per geometric unit cell (namely square and
triangular)  one may therefore only have
two or three points to extrapolate to the infinite-lattice limit \cite{ed_archimedean}.
We mention, that due to the sign problem\cite{TrWi05} frustrated quantum
magnets cannot be treated adequately
by efficient Quantum Monte Carlo (QMC) techniques. 
Hence,  a clear picture regarding
the existence of GS magnetic long-range order (LRO) for some 
of the ALs has yet to emerge.

In this paper we analyze the GS energy $E_g$ and
the magnetic order parameter (sublattice magnetization) $M$ of the HAFM on all of the 
ALs for the extreme quantum case, i.e. spin quantum number  $s =1/2$, by using the coupled-cluster method 
(CCM). The corresponding Hamiltonian is given by
\begin{equation}
H = J \sum_{\langle i,j \rangle } {\bf s}_i \cdot {\bf s}_j \;.
\label{eq2}
\end{equation}
The symbol $\langle i,j \rangle$ indicates those bonds connecting 
NN sites (counting each bond once only) on all of the ALs. We set the energy
scale by putting $J=1$. 

We illustrate here only some relevant features of
the CCM.
For more general information on the methodology of the CCM, see, e.g.,
Refs.~\onlinecite{ccm_theory,zeng98,bishop98a,bishop04}.
CCM 
has recently been applied computationally at high orders of approximation 
to quantum magnetic systems with much success, see, e.g., 
Refs.~\cite{spin_systems_book,spin_half_xxz,ccm_j_prime,ccm_shastry,ccm_extra,ccm_maple,kagome_general_s}.  In the field of quantum magnetism, 
advantages of this approach are that it can be applied 
to strongly frustrated quantum spin systems in any dimension and 
with arbitrary spin quantum numbers.

A basic element of the single-reference CCM used here is the  parameterization the ket
GS eigenvector 
$|\Psi\rangle$  of a general 
many-body system described by a Hamiltonian $H$
(where $H |\Psi\rangle = E_g |\Psi\rangle$)
via $ |\Psi\rangle = {\rm e}^S |\Phi\rangle$ and 
where $S=\sum_{I \neq 0} {\cal S}_I C_I^{+}$. For spin systems the model or reference
state
$ |\Phi\rangle$ is related to the classical GS and the many-body creation
operators $C_I^+$ applied to 
$ |\Phi\rangle$ can be expressed by appropriate products of spin-flip
operators
\cite{spin_systems_book,spin_half_xxz,ccm_j_prime,ccm_shastry,ccm_extra,ccm_maple,kagome_general_s}.
For the unfrustrated ``bipartite'' lattices (namely, square, CaVO, 
SHD, and honeycomb), the model state 
$|\Phi\rangle$ is taken to be the classical collinear two-sublattice N\'eel 
GS. 
For the frustrated ``non-bipartite'' lattices (namely, triangular, 
kagome, star, maple-leaf, trellis, SrCuBO, and bounce), non-collinear
classical GSs are typical. An exception is the SrCuBO lattice, which has a pattern of
exchange bonds that is topologically equivalent\cite{ed_archimedean} 
to the famous Shastry-Sutherland
model \cite{Shastry}. For this frustrated model also the collinear two-sublattice
N\'eel ground state is appropriate as our model
state \cite{ed_archimedean,Mila,ccm_shastry}.
For the 
triangular lattice we have the well-known 120$^{\circ}$ three-sublattice
state.  For the maple-leaf and bounce lattices the classical GS used as
model state has six
sublattices with a characteristic pitch angle \cite{ed_archimedean,ccm_maple}. 
The classical GS of the trellis lattice is an incommensurate spiral one along a
chain \cite{ed_archimedean,trellis}. As quantum fluctuations may lead to a
``quantum'' pitch angle that deviates from the classical
one \cite{ccm_j_prime,ccm_shastry}, 
we consider the pitch angle in the
model states of the maple-leaf, bounce and trellis lattices as a free parameter.  
The case for the kagome and star lattices is
more subtle as there are an infinite number of possible
classical ground states to choose from. However, 
current understanding is that quantum fluctuations 
favor of coplanar states for these systems, such as 
$\sqrt{3}\times\sqrt{3}$ and $q=0$
states \cite{ed_archimedean,chub92,henley1995,sachdev1992},
which are used here as model states.

To perform the CCM calculations for quantum many-body problems one has naturally to use approximations. 
Here we utilize the LSUB$m$  
approximation scheme, in which all $m$-body clusters
spanning a range of no more than $m$ adjacent lattice 
sites are retained (for details, see
Refs.~\onlinecite{spin_systems_book,spin_half_xxz,ccm_j_prime,ccm_shastry,ccm_extra,ccm_maple,kagome_general_s}). To analyze the GS magnetic LRO we consider 
the sublattice magnetization $M(m)$ that can be straightforwardly calculated
within a certain CCM-LSUB$m$
approximation \cite{spin_systems_book,ccm_j_prime,ccm_shastry}.
For more information about the definition of the order parameter $m^+$ used
in the ED study of the ALs in Ref.~\onlinecite{ed_archimedean}, see pages 93 to 94 of that reference.

\begin{table} 
\caption{Extrapolated CCM results for the GS energy per bond of the spin-1/2 HAFM on the various Archimedean 
lattices compared to  ED results from Ref. \cite{ed_archimedean} and other
available data. (Results for the star
and kagome lattices are given for the $q=0$ and $\sqrt{3}\times\sqrt{3}$ model
states.) }
\begin{tabular}{|c|c|c|c|} \hline
Lattice 		& CCM     & ED (Ref.\onlinecite{ed_archimedean})	   &
other results \\  \hline
{\it Bipartite}	&	                  		&	         		& 			 	\\ 
square 		&$-0.3348$			&$-$0.3350 	& $-$0.3347
[\onlinecite{sandvik97}] \\ 
honeycomb 	&$-0.3631$			&$-$0.3632 	& $-$0.3630
[\onlinecite{reger89}]   \\ 
CaVO 	        &$-0.3689$ 			&$-$0.3689	& $-$0.3691
[\onlinecite{spin_half_cavo_II}]   \\ 
SHD 		&$-0.3702$ 		 	&$-$0.3713 	& $-$0.3688
[\onlinecite{shd}]   \\   \hline
{\it Frustrated}	&	                  		& 
	                   & 			 	 \\
SrCuBO 	         &$-0.2312$		 	&$-$0.2310	&
 $-0.23 \ldots 0.24$  [\onlinecite{Mila2013}] \\ 
triangular	 &$-0.1843$	               &$-$0.1842 	& $-$0.1823
[\onlinecite{zhito2009}]    \\ 
{bounce} 	 &$-0.2824$		       &$-$0.2837 	&  \\ 
{trellis}        &$-0.2416$	 	       &$-$0.2471 	&     \\ 
{maple-leaf} 	 &$-0.2124$		       &$-$0.2171 	&    \\ 
kagome       &                              & $-0.2172$         &   $-0.2190
\ldots 0.2193$ [\onlinecite{Yan2011,scholl}] \\      
$q=0$            &$-$0.2179	               & 	&    \\ 
$\sqrt{3}\times\sqrt{3}$ &$-$0.2159	 	& 	&     \\ 
star          &	                                &$-0.3093$  & $-0.316 \ldots 0.318$
[\onlinecite{starIII}]	    \\
$q=0$   	         &$-$0.3110	               & 	  &  \\ 
$\sqrt{3}\times\sqrt{3}$   	         &$-$0.3101    &      &   \\ \hline
\end{tabular} 
\label{tab1} 
\end{table}

\begin{table} 
\caption{
Extrapolated CCM results for the order parameter $M$ 
(of the spin-1/2 HAFM on the various Archimedean 
lattices compared to  ED results from Ref. \cite{ed_archimedean} and other
available data. (Results for the star
and kagome lattices are given for the $q=0$ and $\sqrt{3}\times\sqrt{3}$
model
states.
) }
\begin{tabular}{|c|c|c|c|} \hline
Lattice 		& CCM     & ED (Ref.\onlinecite{ed_archimedean})	   &
other results \\  \hline
{\it Bipartite}	&	                  		&	         		& 			 	\\ 
square 				&0.619		 	&0.635  &
0.614\ldots0.617 [\onlinecite{sandvik97,wiese1996}] \\ 
honeycomb 			&0.547		 	&0.558    & 0.535
[\onlinecite{spin_half_honeycomb}] \\
CaVO 	         		&0.431			&0.461   &  0.356
[\onlinecite{spin_half_cavo_I}] \\ 
SHD 		 	 	&0.366	 	 	&0.425  & 0.509
[\onlinecite{shd}]  \\   \hline
{\it Frustrated}	&	                  		&	                   & 	 	 \\
SrCuBO 	        		&0.404	 	&0.456   & 0.42
[\onlinecite{Mila2013}] \\ 
triangular		 	&0.373          &0.386  & 0.410
[\onlinecite{dmrg_trian}]  \\ 
{bounce} 	         	&$M_I$: 0.122		 	&0.286 &   \\ 
 	                      		&$M_{II}$: $0$		& 	&   \\ 
{trellis} 			 	&$M_I$: 0.040	 	&0.222   &  \\ 
      			 	       &$M_{II}$: $0$	& 	&    \\ 
{maple-leaf} 			&$M_I$: 0.178		 	&0.218  &  \\ 
 	         		      &$M_{II}$: $0$ 		& 	&   \\ 
kagome       &                       &              	&0   \\      
$q=0$   &$0$ 	                & 	&    \\ 
$\sqrt{3}\times\sqrt{3}$ 	 	&$0$ 	& 	&     \\ 
star          &	                                     	&0.094...0.15   & \\
$q=0$   	         	 	&$0$                & 	  &  \\ 
$\sqrt{3}\times\sqrt{3}$   	        	 	&$0$                &&   \\ \hline
\end{tabular} 
\label{tab2} 
\end{table}

Since the LSUB$m$ approximation becomes exact only in the
limit
$m \rightarrow \infty$, it is useful to extrapolate the LSUB$m$ results
in this limit.
For the GS energy the extrapolation scheme
$E_g(m)/N = E_g(m=\infty)/N  + a_1/m^2 + a_2/m^4$ is
well-established\cite{spin_systems_book,spin_half_xxz,ccm_j_prime,ccm_shastry,ccm_extra,ccm_maple,kagome_general_s}.
For the magnetic order parameter $M$ the choice of an appropriate extrapolation
scheme is more subtle.
In cases where GS magnetic LRO is present, e.g. for the square
lattice, the scheme I with $M(m) = M_{\rm I}(m=\infty) + b_1/m + b_2/m^2$ leads to excellent
results for the order parameter\cite{spin_systems_book,spin_half_xxz}. 
On the other hand, for systems where the GS magnetic LRO
is unstable,  the scheme II with $M(m) = M_{\rm II}(m=\infty) + c_1/m^{1/2} + c_2/m^{3/2}$ is
favorable\cite{ccm_extra,kagome_general_s}.
It is also well-known that low-level  LSUB$m$ approximations are 
poor approximations, and they do not    
follow the extrapolation rules well. Hence,
LSUB2 and  LSUB3 data are excluded from extrapolation.
Moreover, since for collinear model states (i.e. for bipartite square,
honeycomb, CaVO, SHD lattices and for the SrCuBO
lattice) no odd-numbered spin flips
appear\cite{spin_systems_book,spin_half_xxz}, we take into account in the
extrapolation for these
lattices only LSUB$m$ data with even $m \ge 4$. On the other hand, for the
triangular, 
kagome, star, maple-leaf, trellis, and bounce lattices where we use     
non-collinear model states (i.e. 
odd-numbered spin flips appear) we take into account in the
extrapolation all LSUB$m$, $m \ge 4$ \cite{comment}.
Due to the different complexity of the lattices and the corresponding model
states the maximum level $m_{\rm max}$ of LSUB$m$
approximations accessible within our CCM code is not unique. Thus, we have
$m_{\rm max}=12$ for the square, honeycomb, CaVO, SHD, $m_{\rm max}=10$ for the 
triangular, kagome, star, SrCuBO, and  $m_{\rm max}=8$ for the bounce,  maple-leaf, and trellis
lattices.

To decide, which extrapolation for the order parameter is appropriate we
proceed as follows:
{First we apply 
both extrapolation schemes I and II.
In case that both schemes lead to $M_{\rm I}(m=\infty)>0$ and $M_{\rm
II}(m=\infty)>0$ we have evidence for GS magnetic LRO, and we use scheme I
for further consideration.
In case that both schemes lead to vanishing $M_{\rm I}(m=\infty)$ and $M_{\rm
II}(m=\infty)$, we have clear evidence for the breakdown of GS magnetic LRO.
However, there are also some cases, where  $M_{\rm I}(m=\infty)>0$ but
$M_{\rm II}(m=\infty)$ vanishes, see table~\ref{tab2}. 
Although,  in these cases a clear statement about GS magnetic
LRO is problematic the magnetic LRO is at least very
fragile, and a non-magnetic cooperative quantum paramagnetic GS is likely.}

It is appropriate to mention earlier attempts to calculate the GS quantities
by means of the CCM for some of the ALs, namely
Refs.~\onlinecite{spin_half_xxz,ccm_square} (square),
Ref.~\onlinecite{ccm_trian} (triangular), Ref.~\onlinecite{ccm_j_prime}
(honeycomb), Ref.~\onlinecite{ccm_cavo}
(CaVO), Ref.~\onlinecite{kagome_general_s} (kagome),
Ref.~\onlinecite{ccm_maple} (maple-leaf), and
Ref.~\onlinecite{ccm_maple} (bounce).
However, most of these previous calculations are limited to lower levels of
approximation LSUB$m$. Hence the new data using higher LSUB$m$ presented here
may yield much more accurate results.

We collect our CCM results for E in Table~\ref{tab1} and for M in
Table~\ref{tab2}. These are compared to those results 
for the ground-state energies  and order parameters $M$ quoted
on page 118 of Ref. \onlinecite{ed_archimedean}. Moreover we also present
available data from previous investigations using other methods.

All the bipartite ALs (and so for unfrustrated HAFM 
systems) exhibit magnetic LRO, where the order parameter is significantly
reduced by quantum fluctuations. This reduction is strongest 
for the two lattices with non-equivalent NN (CaVO and SHD), indicating a
possible instability  against a non-magnetic valence-bond
state \cite{spin_half_cavo_I}.
Thus the order
parameter for the SHD lattice is only $37$\% of the classical value. 
Note that our data for the bipartite square, honeycomb and CaVO lattices are in good agreement with
available QMC
data\cite{sandvik97,wiese1996,spin_half_honeycomb,spin_half_cavo_I,spin_half_cavo_II},
which can be considered as benchmark results. 
For the  bipartite SHD lattice no QMC data are published, the results
reported in Ref.~\onlinecite{shd} are obtained by a variational technique
that might be less accurate than our high-order CCM results.  

For the frustrated lattices the QMC cannot serve as benchmark approach.
Hence, typically the previously published results may have limited
accuracy and our high-order CCM data contribute to a refinement of the GS
data and a better
understanding of these frustrated quantum HAFMs.  
The reference data quoted in Tables~\ref{tab1} and \ref{tab2} are obtained
by tensor-network approach\cite{Mila2013}, spin-wave theory\cite{zhito2009},
 density-matrix renormalization group
 method\cite{Yan2011,scholl,dmrg_trian}, and
bond-operator technique\cite{starIII}.
Among the frustrated ALs the SrCuBO lattice is special, since it is the only
lattice having a  classical collinear N\'eel GS.
Hence it is not surprising that the quantum GS possesses N\'eel LRO
with the largest order parameter $M$ of the frustrated ALs.  
However, the effect of frustration is obvious by a noticeably reduced  $M$
compared to the square lattice.

Particular attention has been paid in the literature to the famous kagome
HAFM \cite{mendels,Yan2011,lauchli2011,scholl,kagome_general_s,normand2013}. 
The good agreement of the CCM GS energy given in Table~\ref{tab1} with recent  
large-scale density-matrix renormalization group\cite{Yan2011,scholl}, 
exact-diagonalization\cite{lauchli2011}, and tensor-network
approach\cite{normand2013} results gives an indication of the accuracy
of our CCM approach for frustrated lattices.
Another example for a non-magnetic GS, first mentioned in Ref.~\onlinecite{ed_archimedean}, 
is the star-lattice HAFM.
For the kagome and the star lattices the two extrapolation schemes
yield vanishing order parameters for both model states, $q=0$ and 
$\sqrt{3}\times\sqrt{3}$, that is consistent with previous studies    
of these
lattices \cite{ed_archimedean,Yan2011,lauchli2011,scholl,kagome_general_s,starI,starII,starIII}.
We mention that for both lattices the CCM GS energy for the $q=0$ model
state is lower than that for the $\sqrt{3}\times\sqrt{3}$ model state.
That is different from previous studies of the GS selection  based on an expansion around the
classical limit\cite{chub92,sachdev1992,henley1995}, where for the kagome
lattice  the
 $\sqrt{3}\times\sqrt{3}$ state
was found to be selected by quantum fluctuations. This may be related  to
the extreme quantum case of $s=1/2$ considered here that is not well
described by  an expansion around the             
classical limit, see also the discussion in
Ref.~\onlinecite{kagome_general_s}.  

For the trellis, maple-leaf and bounce lattices the results are
less clear, since both extrapolation schemes lead to different conclusions
with respect to magnetic LRO. In general, for systems near to a quantum
critical point the results may depend on details of both 
the extrapolation scheme and orders of approximation
used \cite{commentII}.
Due to the computational difficulty for these lattices,
these extrapolations used LSUB4 to LSUB8 approximations only, 
the lowest orders of approximation used in this paper, and we 
find that extrapolation scheme I yields a small but finite  order parameter 
$M$ that is significantly below the ED results reported in
Ref.~\cite{ed_archimedean}. 
By way of comparison, we note that extrapolations using 
scheme 1 with LSUB5 to LSUB8 indicate that the order parameter is 
29\% , 21\%, and 29\% of its classical value for the bounce, trellis, 
and maple-leaf lattices.
Note also that the finite-size extrapolation
of the ED data for these lattices is particularly poor, since only two
(bounce, maple-leaf) or 
three (trellis) data points could be used.   Moreover, periodic boundary
conditions used for ED calculations in Ref.~\cite{ed_archimedean} might be  not
well-suited for the incommensurate
spiral correlations present for the trellis lattice.
On the other hand,  extrapolation scheme II using LSUB4 to LSUB8 
for these three lattices lead to vanishing order parameters. 
Hence, we may conclude that these lattices exhibit either a weak
magnetic LRO or are even in a magnetically disorder GS phase. Hence, 
they are candidates to find non-magnetic states in experiments, see also
Ref.\cite{fennell}. 
We mention that in Ref.~\onlinecite{trellis} by means
of spin-wave and variational techniques similar conclusions for the trellis
lattice were found.

\begin{figure}[ht]
\epsfxsize=13cm
\centerline{\epsffile{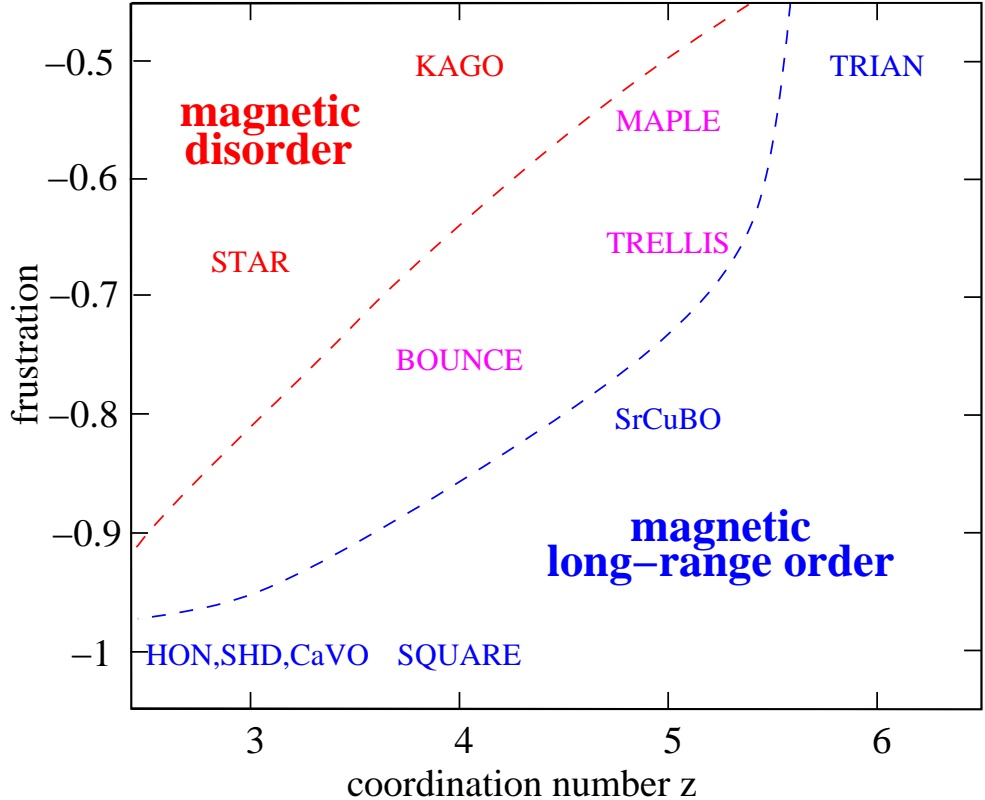}}
\caption{Sketch of semi-classical magnetic order and quantum magnetic
disorder of the ALs in a parameter space spanned by 
 frustration (classical GS energy per bond, see
 Ref.~\onlinecite{frustration})  and 
coordination number $z$. 
}
\label{fig2}
\end{figure}

In this manuscript we have presented a survey of results for ground-state energy and
order parameter of the $s=1/2$ Heisenberg antiferromagnet on all 
eleven Archimedean lattices by using the CCM.  In 2D quantum magnets the
competition
between fluctuations  and interaction determines the GS features. 
Our results show a clear correspondence between lattice topology  and
existence of  GS magnetic LRO. The most important ingredients 
affecting the magnetic ordering are geometric
frustration and the coordination number. Moreover, the competition of
non-equivalent NN bonds is relevant.
To illustrate the  role of geometric             
frustration and coordination number we summarize our findings  
in Fig.~\ref{fig2} in a parameter space spanned by 
 frustration\cite{frustration}  and
coordination number $z$.
Clearly there are three regions of magnetic GS ordering: semiclassical magnetic LRO
(collinear or non-collinear), magnetic disorder (cooperative quantum paramagnetism) and
an intermediate  region with ALs, namely trellis, bounce, maple-leaf, which
may have either  a GS with fragile magnetic LRO, a critical GS order or a
GS with weak disorder.  
This group of ALs deserves particular further attention to clarify the
nature of the GSs.  
We think that our results ought to 
provide also a useful benchmark for the Archimedean lattices to which experimental 
studies and other approximate theoretical methods might be tested. 



\begin{thebibliography}{200}
\bibitem{comment} The same rule how to perform the extrapolation was used in our previous paper
on the kagome lattice.\cite{kagome_general_s}
\end{thebibliography}
\end{document}